\documentclass[aps,prl,twocolumn,showpacs,superscriptaddress]{revtex4}
\usepackage{amsmath}
\usepackage{dcolumn}
\usepackage{epsfig}
\usepackage{graphicx}
\usepackage{latexsym}
\usepackage{amssymb}
\def\S2RO4{Sr$_2$RuO$_{4}$}
\def\SRO3{SrRuO$_{3}$}
\def\STO3{SrTiO$_{3}$}

\begin{document}

\title{Critical thickness for itinerant ferromagnetism in ultrathin films of \SRO3 }

\author{Jing Xia}
\affiliation{Geballe Laboratory for Advanced Materials, Stanford University, Stanford, California, 94305}
\affiliation{Department of Physics, Stanford University, Stanford, CA 94305} 
\author{W. Siemons}
\affiliation{Geballe Laboratory for Advanced Materials, Stanford University, Stanford, California, 94305}
\affiliation{Faculty of Science and Technology and MESA+ Institute for Nanotechnology, University of Twente, P.O. Box 217, 7500 AE, Enschede, The Netherlands}
\author{G. Koster}
\affiliation{Geballe Laboratory for Advanced Materials, Stanford University, Stanford, California, 94305}
\affiliation{Faculty of Science and Technology and MESA+ Institute for Nanotechnology, University of Twente, P.O. Box 217, 7500 AE, Enschede, The Netherlands}
\author{M.R. Beasley}
\affiliation{Geballe Laboratory for Advanced Materials, Stanford University, Stanford, California, 94305}
\affiliation{Department of Applied Physics, Stanford University, Stanford, CA 94305} 
\author{A. Kapitulnik}
\affiliation{Geballe Laboratory for Advanced Materials, Stanford University, Stanford, California, 94305}
\affiliation{Department of Physics, Stanford University, Stanford, CA 94305}
\affiliation{Department of Applied Physics, Stanford University, Stanford, CA 94305}

\begin{abstract}
Ultrathin films of the itinerant ferromagnet  \SRO3 were studied using transport and magnto-optic polar Kerr effect. We find that below 4 monolayers the films become insulating and their magnetic character changes as they loose their simple ferromagnetic behavior.  We observe a strong reduction in the magnetic moment which for 3 monolayers and below lies in the plane of the film. Exchange-bias behavior is observed below the critical thickness, and may point to induced antiferromagnetism in contact with ferromagnetic regions. 
\end{abstract}


\maketitle

\SRO3 is an itinerant ferromagnet with an orthorhombically distorted cubic perovskite structure, exhibiting a transition to a ferromagnetic state at $T_c \sim$160 K that was shown to be dominated by transverse fluctuations of robust local moments  of size $\sim 1.6 \mu_B$ \cite{dodge}, the largest of any 4-d ferromagnet.  While at high-temperatures, in the paramagnetic phase, it exhibits  a ``bad metal" behavior in the limit of $k_F\ell \sim 1$ \cite{klein1} suggesting that  Fermi liquid theory may not be valid, the observation of quantum oscillations in the electrical resistivity of high-quality thin films of  \SRO3 demonstrated that the ground state of this system is a Fermi liquid \cite{mackenzie}.  At the same time, the degree of electron correlation in \SRO3 has been found to be a strong function of ruthenium deficiency \cite{wolter1}.  To understand the contrast in the behavior of \SRO3 between high and low temperatures, appropriate perturbations such as disorder and reduced dimensionality, may be used that directly disturb the magnetic and transport properties of the system. Indeed, recent studies of the thickness dependence of the transport and electronic structure of \SRO3 films \cite{herranz1,Toyota} concluded that a metal-insulator transition (MIT) occurs in these films at a critical film thickness of 4 or 5 monolayers (ML), depending on disorder.  However, the reported island-like microstructure showing coalescence of three-dimensional patches, and the inability to study the nature of the magnetism, hinder any possible understanding of the observed transition.  Since this may be the first example of the interplay between itineracy, ferromagnetism disorder and dimensionality, better films growth and a more direct probe of magnetism are needed to establish the important ingredients of the physics involved.

In this paper we present new results on the MIT in ultrathin \SRO3 films and their associated magnetic properties. We show that in homogeneous films of \SRO3 a metal-insulator transition (MIT) occurs at a critical thickness below 4 ML.  While $T_c$ drops rapidly below  $\sim$10 ML, the size of the moment remains unchanged from its $1.6\mu_B$ in thick films \cite{dodge}, and the easy axis which has been closer to normal for thick films, becomes even more normal.  However, below the critical thickness the easy axis of the moment plummets to the plane of the film, and an exchange-bias behavior emerges, suggesting  the existence of antiferromagnetic (AFM) regions in the different layers that interact with the remaining ferromagnetic regions. Examination of the transport properties of the measured films shows an increase in the resistance  with decreasing thickness. At 4 ML the extrapolated low-temperature sheet-resistance is of order $\sim$7 k$\Omega$, jumping up 8 orders of magnitude in 3 ML films. 

\SRO3 samples used in our experiment were grown by Pulsed Laser Deposition (PLD). The samples were grown in a vacuum chamber with a background pressure of 10$^{-7}$ Torr.  A 248 nm wavelength KrF excimer laser was employed with typical pulse lengths of 20-30 ns. The energy density on the target is kept at approximately 2.1 J/cm$^{2}$.  All films were grown on TiO$_2$ terminated \STO3 substrates \cite {Koster}, at 700 C, with a a laser repetition rate of 4 Hertz. We have calibrated the deposition rate multiple times throughout the process by performing x-ray reflectivity on thicker samples. The thickness of the films range from 2 to 25 ML, each with uncertainty of only few laser pulses, which is equivalent to a very small fraction of a 1 ML (approx. 20 pulses per 1 ML). 

\begin{figure}[h]
\includegraphics[width=1.0 \columnwidth]{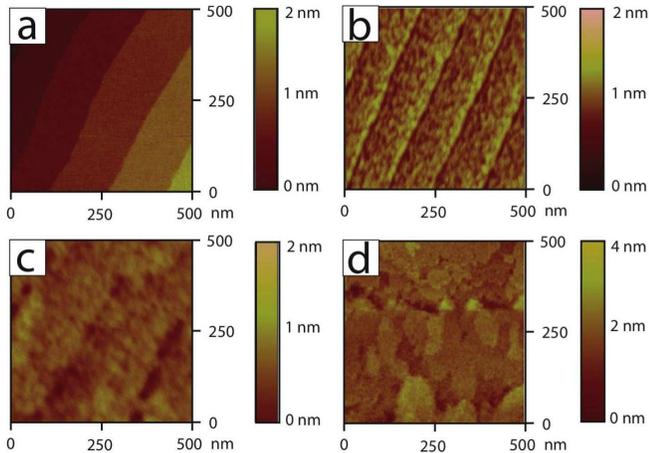}
\caption{AFM images of:  (a) \STO3 substrate before deposition,  (b) 2 ML, (c) 5 ML and (d) 9 ML (see text).}
\label{AFM}
\end{figure}

Atomic force microscope (AFM) images (Fig.~\ref{AFM}), taken immediately following  deposition,  indicate that between 2 and 7 ML, \SRO3 films show homogeneous coverage of the substrate, with two-dimensional stripe shaped steps following the ($\sim 0.2 ^{\circ}$) miscut of the substrate. These two-dimensional steps are 1 ML in height and are typically 100 nm in width. Moreover, unlike previous reports \cite {Herranz,Toyota}, no three-dimensional island growth was observed, indicating an atomically smooth surface and single domain structure in these films. The observed step-like growth seems to fade at thicknesses above 9 ML as the steps  mostly  coalesce, suggesting a transition from a growth mode of  two-dimensional layer-by-layer  to a step flow mode, in agreement with earlier reports \cite {Choi}.

\begin{figure}[h]
\includegraphics[width=1.0 \columnwidth]{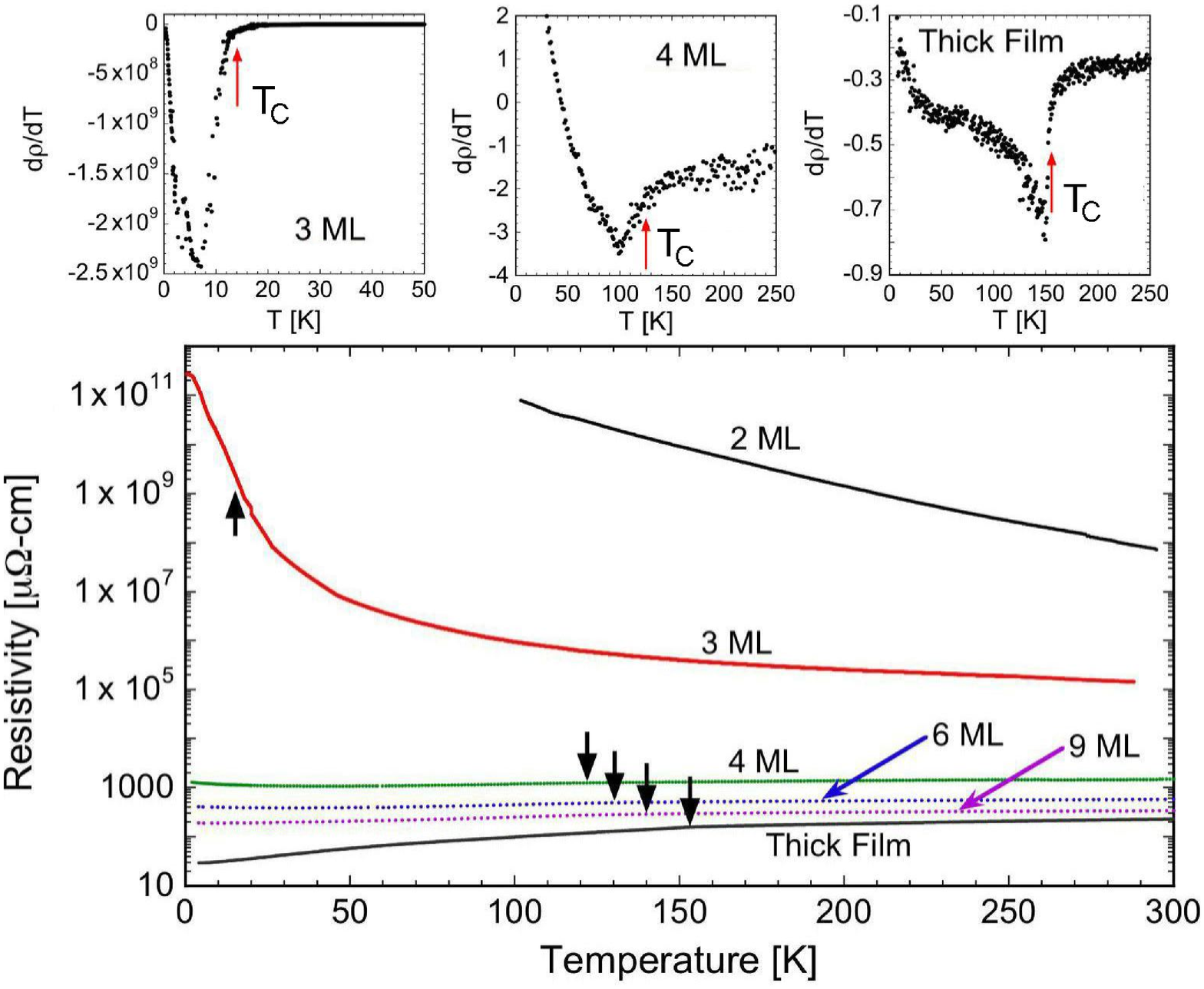}
\caption{Resistivity data of \SRO3 films.  Arrows point to the location of the ferromagnetic transition, determined from the derivative of the resistivity (see e.g. top panels for 3 ML, 4 ML and thick film derivatives).}
\label{rvt}
\end{figure}

Fig.~\ref{rvt} shows the resistivity of the films through the transition, measured from room temperature down to  4.2K. The ferromagnetic transition was noticeable in all films of 4 ML and above, however the transition becomes broad and difficult to determine for the very thin films.  We note that while  the extrapolated low-temperature sheet-resistance of the 4 ML film is of order $\sim$7 k$\Omega$, the low-temperature resistance of the 3 ML film increases  more than 8 orders of magnitude, much higher than the quantum of resistance for two-dimensions of $h/e^2\sim$26 k$\Omega$. Thus it is clear that a metal-insulator transition has occurred in between these two thicknesses.

The magnetic properties of the films were determined from Polar Kerr effect (PKE) measurements, which is only sensitive to the out-of-plane component of the magnetization \cite{bader}.  While in general for thin-films magnetism the Kerr signal is large \cite{bader},  for ultrathin films (approaching 1 ML) of weak ferromagnets, especially deposited on strongly birefringent substrates, these measurements may become difficult because of a small signal that may be masked by the rotation of polarization due to the substrate. In our case, \SRO3 films are deposited on miscut substrates of \STO3 which are very strongly linearly birefringent. To overcome the above difficulties we have used a zero-area-loop Sagnac interferometer \cite{Xia2}, which directly detects  the circular birefringence in the magnetized sample. This unique design is based on a Sagnac loop in which two counter-propagating beams  with opposite circular polarization reflect from the sample while completing a Sagnac loop.  This design which was first introduced by Xia {\it et al.} \cite{Xia2} is capable of measuring time-reversal-symmetry breaking effects with a shot-noise limited sensitivity of 100 nanorad/$\sqrt{Hz}$ at a power of 10 $\mu$Watt, while being completely immune to any reciprocal effects in the sample, such as linear birefringence \cite{Xia}.  For the results reported in this paper we used a normal-incidence configuration, measuring at a wavelength of 1550 nm with a beam focused on the sample to a spot size of 3 $\mu$m and in the temperature range of 0.3 K to room temperature.  Since the optical penetration depth at the used frequency is of order 200 nm, while the thickest sample used was only 8.8 nm, the signal measured, to a very good approximation, was simply proportional to the area density of the magnetic moment\cite {bader}.

\begin{figure}[h]
\includegraphics[width=1.0 \columnwidth]{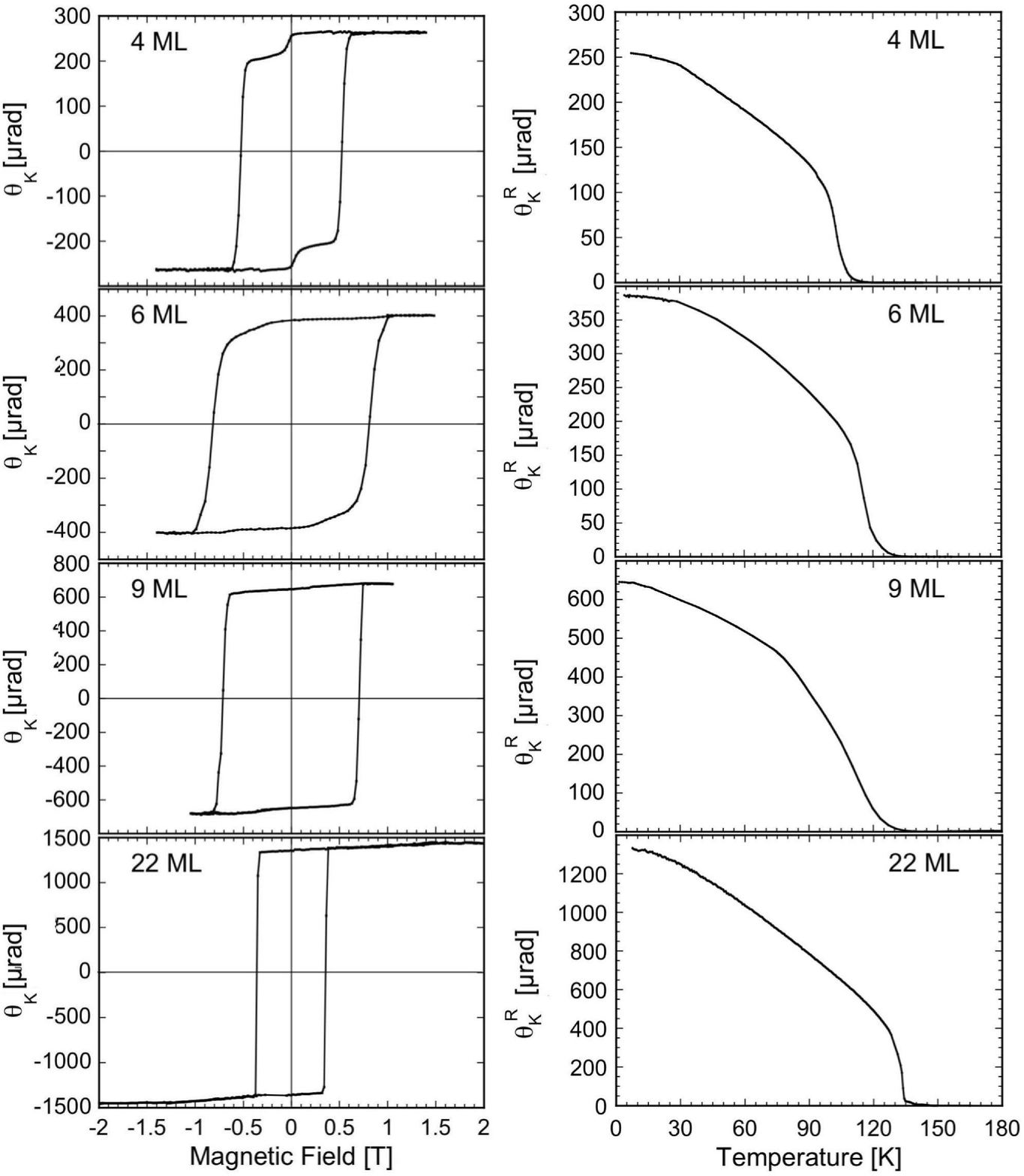}
\caption{ Panels a-d are PKE Hysteresis loop for \SRO3 films of different thickness taken at 4 K, with magnetic field applied perpendicular to the plane of the film.  panels e-h show the temperature dependence  remanent PKE signal measured at zero magnetic field during warmup, after a positive saturation magnetic field was turned off at the lowest temperature. }
\label{combined}
\end{figure}

Fig.~\ref{combined} shows the evolution of the PKE measured on the samples from 4 ML  to 22 ML thick samples. Hysteresis loops were obtained by recording the Kerr signal at the lowest temperature (typically 0.4 K) while ramping an out-of-plane magnetic field, and then subtracting the linear paramagnetic response from the \STO3 substrate and diamagnetic response from the optical components in the fringing magnetic field.  After the magnetic field was turned off, the PKE was measured as a function of temperature while the films were warmed to room temperature.  This allowed the determination of the Curie temperature $T_c$  and the angle of the easy axis. 

\begin{figure}[h]
\includegraphics[width=1.0 \columnwidth]{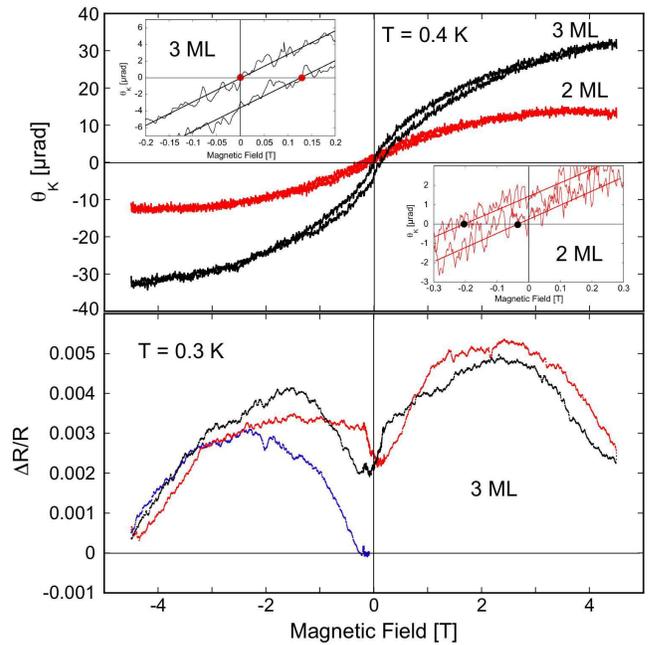}
\caption{ a) Hysteresis loop for the 2 ML and 3 ML samples. Insets show the region near the origin where the exchange-bias nature of the loop is clear. 3 ML is biased to the left and the 2 ML is biased to the right. Thick-dots mark the crossing of the field axis. b) Magnetoresistance for the 3 ML sample. Loop starts at S, continues to A, then B, and ends at A. Subsequent loops trace the A-B-A loop.}
\label{hyst}
\end{figure}

The temperature dependent remanent-Kerr-effect of the 2 and 3 ML films did not show any ferromagnetic transition down to 0.4 K, to a resolution of $\pm$0.2 $\mu$rad. However, we suggest that the magnetic transition may be deduced from the resistivity data that shows sharp upturn in the resistivity of the 3 ML film below $\sim$25 K (note the logarithmic axis!). The resistivity of the 2ML film could not be measured below 100 K due to the large resistance. Of course, this assertion still needs to be checked. Magnetization curves of  2 and 3 ML films at 0.4 K are shown in Fig.~\ref{hyst}a. These were obtained by first cooling in a field, then performing a hysteresis loop, followed by a subtraction of the (diamagnetic) contribution of the fiber strand in the magnet. The S-shape and finite opening of both curves indicate that the low-temperature phase of these films have ferromagnetic component with moments that lie entirely in the plane of the film. The open loops are non-symmetric with respect to zero-field, reminiscent of  exchange-bias behavior \cite{nogues}.  Exchange-bias (EB) phenomena originate at  the interface of FM and AF regions, where uncompensated AF moments result in a bias magnetic field, causing the hysteresis loop of the FM to be shifted away from the origin  \cite{nogues}.  Indeed,  the hypotheses of both, remanent in-plane ferromagnetism and  EB are further supported by magnetoresistance (MR) measurements at 0.3 K shown in Fig.~\ref{hyst}b. We note that  the maximum MR observed is $\Delta R/R \sim 0.005$, a very small effect when compared to spin-scattering dominated MIT.

\begin{figure}[h]
\includegraphics[width=1.0 \columnwidth]{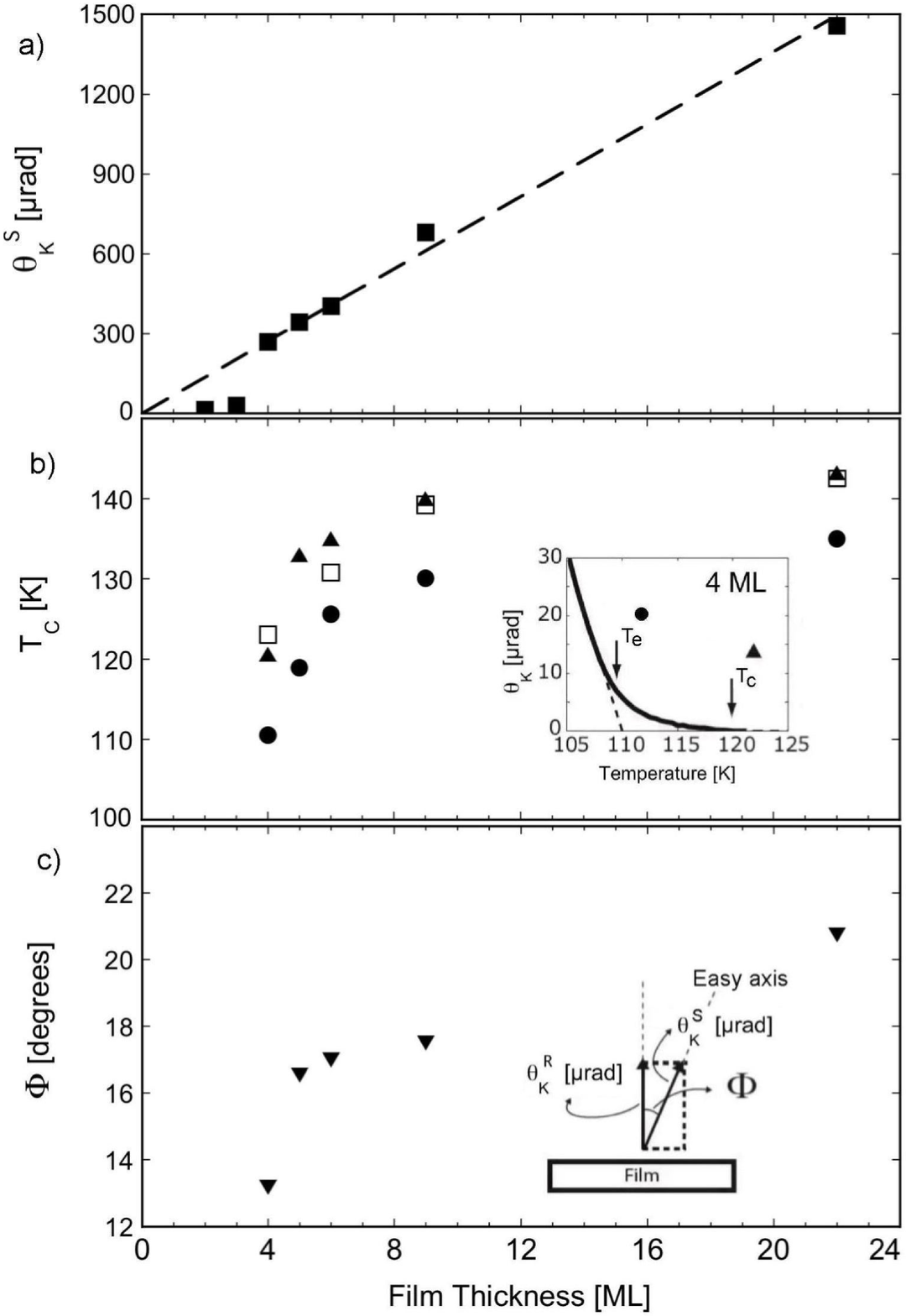}
\caption{ Thickness dependencies of (a) saturated Kerr signal ($\blacksquare$) at the lowest temperature; (b) Curie temperature $T_c$ ($\blacktriangle$), extrapolated Curie temperature $T_e$ ($\bullet$), Resistivity anomaly ($\square$); and (c) the angle $\Phi$ ($\blacktriangledown$) between film normal and magnetic easy axis at the lowest temperature. Dashed line in (a) is the linear fit of the data point between 4 and 22 ML. Insert in (b) shows how $T_c$ and $T_e$ are determined.}
\label{alldata}
\end{figure}

The first sharp hysteresis loop is obtained for the 4 ML sample (Fig.~\ref{combined}a), pointing to ferromagnetism with an almost perpendicular moment \cite{steps}.   Turning off the magnetic field, a remanent signal is observed ($\theta_K^R(T)$), that disappears at  $T_c$ (Fig.~\ref{combined}e).  Similar data for other samples  is given in Fig.~\ref{alldata} where  we show the thickness dependence of the saturation Kerr signal  ($\theta_K^S(T)$, which is determined as the highest point of the hysteresis loop in Fig.~\ref{combined}), the $T_c$ of the layers, and the variation of the easy-axis for all ferromagnetic films.  Fig.~\ref{alldata}a shows that the saturated Kerr signal is proportional to the film thickness from the thick (22 ML) down to the thinnest samples (4 ML), extrapolating to zero thickness. Since we argued that all these films are in the very thin limit compared to the penetration depth of the light, this result clearly shows that the thick film moment ($\sim 1.6 \mu_B$) does not change, and that that all layers are ferromagnetic.  Below 4 ML the saturation Kerr signal plummets, indicating a much smaller moment of  $\sim 0.2 \mu_B$. 

Fig.~\ref{alldata}b shows the thickness dependence of $T_c$. To determine the temperature above which no ferromagnetism is observed we magnified the region near the transition as show in the inset to that figure. While in general the magnetization vanishes at $T_c$ with an exponent smaller than unity, domain structure and reorientation in films may result in apparent lower transition temperature ($T_e$) \cite{gerber1}.  We therefore define two critical temperatures as show in the inset, and plot both in  Fig.~\ref{alldata}b. We note that it is $T_c$, the temperature at which the Kerr signal vanishes, which smoothly extrapolates to the thick films limit and therefore to previously published data on 3-dimensional \SRO3 films \cite{klein1}.  Both, $T_e$ and $T_c$ cannot be measured below 4 ML.  We note that the anomaly in the resistivity, measured by taking the derivative of the resistivity curves,  agrees with $T_c$ for the very thin films and continues towards the thick films limit as expected \cite{klein1}.

To determine the magnetic anisotropy angle,  we calculate $\Phi={\rm cos}^{-1}(\theta_K^R/\theta_K^S)$. This is plotted in  Fig.~\ref{alldata}c. It was previously found that for high-quality epitaxial thick films and at low temperatures $\Phi \sim 30^\circ$ \cite{klein1}.   In Fig.~\ref{alldata}c we show that  $\Phi$ decreases from $22^\circ$ in the case of 22 ML film,  to $14^\circ$ in 4 ML film.  Thus, the fact that below 4 ML the moment is almost entirely in the plane ( Fig.~\ref{hyst}), hence going in the complete opposite way to the trend we found above, is just another confirmation that a phase transition occurred between 3 and 4 monolayers.

The above observations point to a unique phase transition that occurs as a function of thickness. Thick films are itinerant and ferromagnetic with the moment pointing almost entirely in the direction perpendicular to the plane of the films. However, when made thin, the moments seem to plummet into the plane of the films, with AFM layers appearing in contact with FM layers thus inducing exchange-bias behavior. The origin of AFM regions may be a consequence of both, increased disorder and the emerging two-dimensional physics, and the observed thickness-driven phase-transition could either be a natural consequence of enhanced correlations. The strong reduction in the moment, from $\sim 1.6 \mu_B$ to $\sim 0.2 \mu_B$ may further indicate a proximity to a quantum critical point in which coupling to fluctuations causes the reduction in the moment.

Discussions with L. Klein are greatly acknowledged. Fabrication of the Sagnac system was  supported by Stanford's Center for Probing the Nanoscale, NSF NSEC Grant 0425897. Work at Stanford was supported by the Department of Energy Grant  DE-AC02-76SF00515.


\begin{thebibliography}{99}

\bibitem{dodge}
J. S. Dodge {\it et al.}, Phys. Rev. B 60, R6987 (1999).

\bibitem{klein1}
L. Klein, {\it et al.}, Phys. Rev. Lett. 77, 2774 (1996).

\bibitem{mackenzie}
A. P. Mackenzie {\it et al.}, Phys. Rev. B 58, R13318 (1998).

\bibitem{wolter1}
W. Siemons, G. Koster, A. Vailionis,  H. Yamamoto, D.H.A. Blank, and M.R. Beasley,  Phys. Rev. B 76, 075126 (2007).

\bibitem{herranz1}
G. Herranz, B. Mart'nez, J. Fontcuberta, F. S\'{a}nchez, C. Ferrater, M. V. Garc'a-Cuenca, and M. Varela, Phys. Rev. B 67, 174423 (2003).

\bibitem{Toyota}
D. Toyota {\it et al.}, Appl. Phys. Lett. 87, 162508 (2005).

\bibitem{Antognazza}
L.Antognazza, K.Char, T.H.Geballe, L.L.H.King, and A.W.Sleight,  Appl.  Phys. Lett. 63, 1005 (1993).

\bibitem{Feigenson}
Michael Feigenson, James W. Reiner, and Lior Klein,  Phys. Rev. Lett. 98, 247204 (2007).

\bibitem{qiu}
Z. Q. Qiu and S. D. Bader,  Rev. Sci. Inst. 71, 1243 (2000).

\bibitem{Koster}
G. Koster, B.L. Kropman, G. J. H. M. Rijnders, D. H. A. Blank, and H. Rogalla, Appl. Phys.  Lett. 73, 2920 (1998).

\bibitem {Choi}
J. Choi, C. B. Eom, G. Rijnders, H. Rogalla, and D. H. A. Blank, Appl. Phys.  Lett. 79, 1447 (2001).

\bibitem {Herranz}
G. Herranz, B. Martinez, J. Fontcuberta, F. S\'{a}nchez, M. V. Garcia-Cuenca, C. Ferrater, and M. Varela, Appl. Phys. Lett. 82, 85 (2003).

\bibitem {bader}
See e.g. S. D. Bader,  J. Magn. Magn. Mater. 100, 440 (1991).

\bibitem {Klein}
L. Klein, J. S. Dodge, T. H. Geballe,  A. Kapitulnik, A. F. Marshall, L. Antognazza, and K. Char,  Appl. Phys. Lett. 66, 2427 (1995).

\bibitem {Herranz2}
G. Herranz  {\it et al.}, J. Appl. Phys. 97, 10M321 (2005).

\bibitem {Xia2}
Jing Xia,  P. T. Beyersdorf,  M. M. Fejer,  and  A. Kapitulnik, Appl. Phys. Lett.,  89, 062508 (2006).

\bibitem {Xia}
Jing Xia, Y. Maeno, P. T. Beyersdorf, M. M.Fejer,  and A. Kapitulnik, Phys. Rev. Lett. 97, 167002 (2006).

\bibitem{nogues}
For a recent review see e.g. J. Nogu\'{e}s and I.K. Schuller, J. Magn. Magn. Mater. 192, 203 (1999).

\bibitem{steps}
We note that below 9 ML we occasionally observed step structure  which is usually attributed to  competition between domain wall and coherent reversal in ultrathin magnetic films: see e.g. R. A. Hyman  A. Zangwill, and M. D. Stiles, Phys. Rev. B 58, 9276 (1998).

\bibitem{gerber1}
O. Riss, A. Tsukernik, M. Karpovsky and A. Gerber, J. Magn. Magn. Mat 298, 73 (2006).

\end{thebibliography}
\end{document}